\documentclass[ aip, amsmath,amssymb,preprint]{revtex4-1}
\usepackage[utf8]{inputenc}
\usepackage{siunitx}
\usepackage{graphicx}
\usepackage{amsfonts}
\usepackage{hyperref}
\usepackage{color}
\usepackage[normalem]{ulem}

\definecolor{orange}{rgb}{1,0.5,0}

\newcommand{\Fs}{F_{sq}}
\newcommand{\Fe}{F_\mathrm{elec}}
\newcommand{\Vs}{V_{sq}} 
\newcommand{\Pdx}{\mathcal{P}(\Delta X_\tau)}
\newcommand{\stdX}{\sigma_{\tau}}

\begin{document}

\title{Calibrated force measurement in Atomic Force Microscopy using the Transient Fluctuation Theorem} 
\author{Samuel Albert}
\email[]{present adress: ISIS, Univ. Strasbourg, France}
\author{Aubin Archambault}
\author{Artyom Petrosyan}
\author{Caroline Crauste-Thibierge}
\email[]{caroline.crauste@ens-lyon.fr}
\author{Ludovic Bellon}
\author{Sergio Ciliberto}
\affiliation{$1$ Univ Lyon, Ens de Lyon, Univ Claude Bernard, CNRS, Laboratoire de Physique, UMR 5672, F-69342 Lyon, France}

\date{\today}

\begin{abstract}
The Transient Fluctuation Theorem is used to calibrate an Atomic Force Microscope by measuring the fluctuations of the work performed by a time dependent force applied between a colloïdal probe and the surface. From this measure one can easily extract the value of the interaction force and the relevant parameters of the cantilever. The results of this analysis are compared with those obtained by standard calibration methods. 
\end{abstract}

\pacs{}

\maketitle 

In the measurement of forces in micro and nano devices, the calibration of the apparatus may be difficult and several techniques are used to improve the accuracy of the instruments\cite{Calib_1,Calib_2,Calib_3,Laurent_2012,paolino2013RSI,de2009no,Bellon_2008}. The purpose of this letter is to show that using stochastic thermodynamics, we can impose on the measured results extra constrains, which may be useful either as an alternative method of calibration or simply as a test. 

Stochastic thermodynamics extends the laws of thermodynamics to small systems where the role of thermal fluctuations cannot be neglected \cite{Ciliberto_PRX,Seifert_2012}. Indeed in these systems not only the mean values of thermodynamic quantities, such as the work, the heat and the entropy, are important but also their fluctuations and their probability density functions (pdf). Many experimental studies have been performed in the recent years to check the theoretical predictions and to use them for several applications \cite{Ciliberto_PRX}. One of the most important results of stochastic thermodynamics is the Transient Fluctuation Theorem (TFT)\cite{EvansFT}, which imposes some general constrains on the pdf of the work performed on a system by external forces. Specifically if the system is in an equilibrium state and a force $F$ is applied at time $t=0$ then the TFT states that the pdf $\mathcal{P}(W_\tau)$ of the work $W_\tau$ performed by $F(t)$ in a time $\tau$ has the following property: 
\begin{equation}
\ln\left(\frac{\mathcal{P}(W_\tau)}{\mathcal{P}(-W_\tau)} \right)= \frac{W_\tau}{k_B T}, \ \ \ \forall \tau
\label{eq:TFT}
\end{equation} 
where $k_B$ is the Boltzmann constant and $T$ the temperature of the heat bath. It is important to notice that for the TFT the system at time $t=0$, when the force $F(t)$ is applied, must be in equilibrium. In this letter we will show how the constrains imposed by TFT can be used to perform calibrated force measurements with an Atomic Force Microscope (AFM) and to check the standard calibration methods. We have applied eq.~\ref{eq:TFT} to the work performed by an external force on an AFM cantilever in a viscous environment. We will show that using eq.~\ref{eq:TFT} we can easily extract the value of the force without knowing the value of the stiffness of the cantilever and the value of the viscous damping. 
\begin{figure} 
	\includegraphics[width=.45\textwidth]{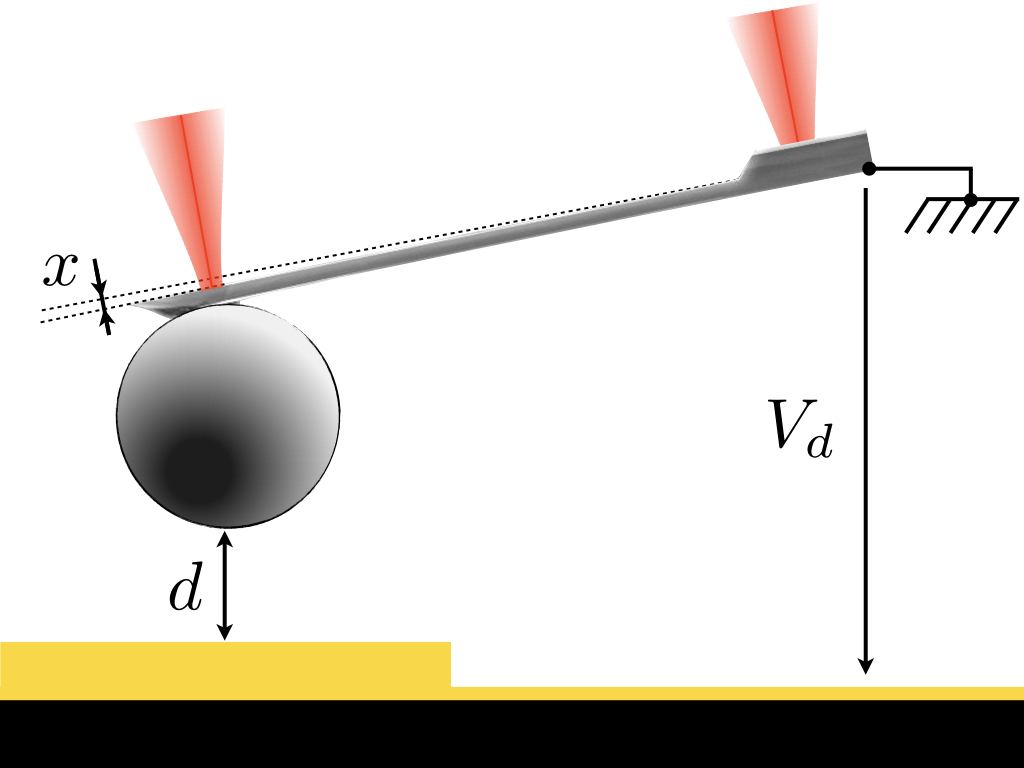}
	\caption{Experimental set-up. A polystyrene bead is glued at the tip of the cantilever using UV cured glue. The bead, the cantilever and the surface are coated with gold, so that a voltage $V_d$ can be used to apply a force. The deflection $x$ is read with a differential interferometer, sketched here by the two laser beams~\cite{paolino2013RSI}.}
	\label{fig:Cellule}
\end{figure}

The experiments are performed using a home made AFM which is characterized by a calibrated measurement and a high resolution: the deflection $x$ of the cantilever is read by a quadrature phase differential interferometer, featuring an intrinsic floor noise of about $10^{-14}\si{\metre}/\sqrt{\si{\hertz}}$ \cite{paolino2013RSI,lecunuder2018}. We use commercial silicon AFM cantilevers (Nanoandmore PPP-ContAu-10) at the tip of which a polystyrene bead is glued. A gold layer is then coated on the sphere/cantilever set to ensure electrical continuity. The cantilever is typically $\SI{450}{\micro\meter}$ long, $\SI{50}{\micro\meter}$ wide and $\SI{2}{\micro\meter}$ thick. The bead radius is $ R = \SI{76 \pm .5}{\micro\meter}$, measured in a SEM before the experiments. 

The cantilever is placed in a cell which can be filled either with a liquid or with nitrogen. The bead is placed above a gold coated glass plate. We use a sphere plane interaction to have a well define geometry that allows us to check the experimental results. A schematic diagram of the set-up is presented in fig.~\ref{fig:Cellule}. A piezoelectric actuator with an integrated displacement sensor allows the control, with an accuracy of $\SI{.2}{\nano\meter}$, of the distance $d$ between the sphere and the plane. The gold coating on both surfaces allows us to apply a voltage difference $V$ between them. This voltage creates an electrostatic attractive force on the bead, which for $d\ll R$ takes the form:
\begin{equation}
 \Fe = \frac{\pi \epsilon_0 R}{d} V^2
 \label{eq:elecforce}
\end{equation}
where $\epsilon_0$ is the vacuum permittivity \cite{Durand_elect,elec_sph_pla}. $\Fe$ can be used as a test force and as a way to measure $d$ by comparing the response of the cantilever to the applied voltage. However one has to take into account that independently from the applied voltage, a contact potential $V_c$ exists between the gold coatings of the bead and the surface. $V_c$ induces an offset in the total voltage which is unknown because it depends on the surface quality~\cite{lecunuder2018,de2009no}. Thus, in eq.~\ref{eq:elecforce}, $V= V_c+V_d$ where $V_d$ is the externally applied voltage and $V_c$ must be experimentally determined. 

Using the piezoelectric actuator, the surface is brought close to the bead, at a distance $d \simeq\SI{2}{\micro\meter}$. At this distance $d \ll R$ and eq.~\ref{eq:elecforce} can be safely applied to determine the interaction force and check the results. 

\begin{figure}[t]
	\centering
	\includegraphics[width =\linewidth]{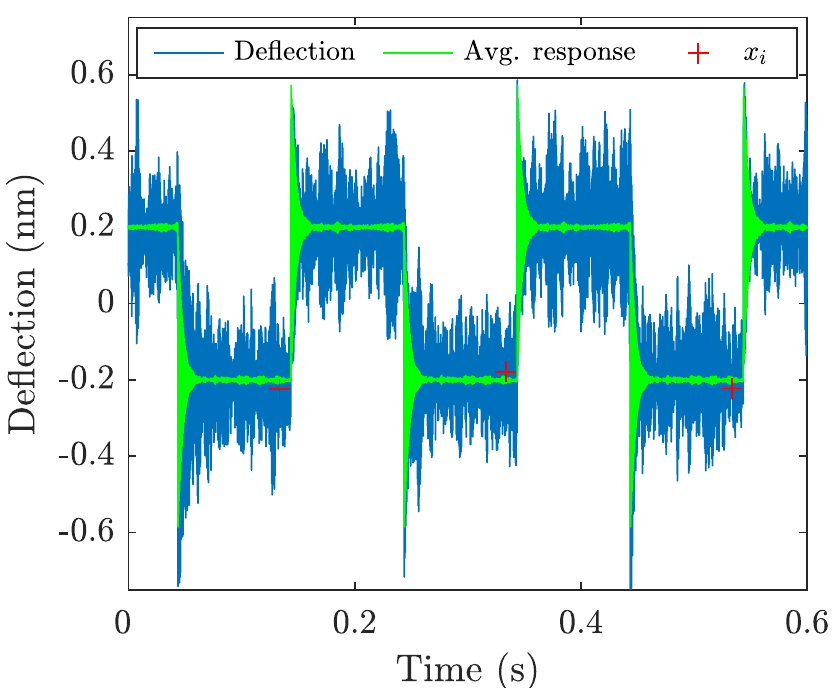}
	\caption{The response of the cantilever driven by a square wave force: deflection of a single experiment (blue), and average over 20 minutes (green). $x_i$ is a reference point taken on the equilibrium state before the square wave rise.}
 	\label{fig:deflection}
\end{figure}

In order to use the TFT for calibration, a square wave voltage between $V_d=\SI{0}{V}$ and $V_d=V_{sq}$ at $\SI{5}{\hertz}$ is applied between the sphere and the plane (see fig. \ref{fig:deflection}). This corresponds to the application of an electrostatic force $\Fe$ of eq.~\ref{eq:elecforce} which periodically changes from $F_i$ at $V=V_c$ to $\Fs$ at $V=\Vs+V_c$. Note that the deflection $x$ is very small ($x\ll d$), so that to a very good approximation $d$ can be considered as a constant during all the protocol, hence the force is also a square wave. Each plateau is much longer than the relaxation time $\tau_\mathrm{relax}\approx\SI{20}{ms}$ of the cantilever to insure that before each step of $V_{sq}$ the cantilever is relaxed to equilibrium. The position $x(t)$ of the cantilever and the applied voltage $V_{sq}(t)$ are sampled for about 20 minutes at $\SI{50}{kS/s}$. 

\begin{figure}[tb]
	\centering
	\includegraphics[width=\linewidth]{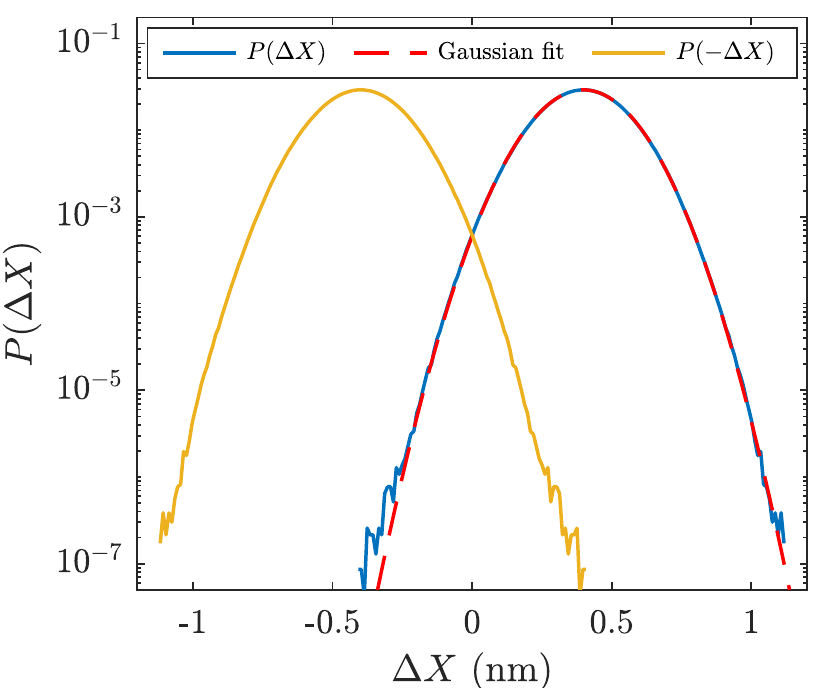}
	\caption{The distribution of $\mathcal{P}(\Delta X)$ and a Gaussian fit, measured at $V_{sq} = \SI{250}{\milli\volt}$. The histogram is computed using the values of all $\tau>2 \tau_\mathrm{relax}$.} 
	\label{fig:Gaussienne}
\end{figure}

Using these data, the value of the force jump $F=\Fs-F_i$ can be measured using eq.~\ref{eq:TFT} without any knowledge of the contact potentials $V_c$, the distance $d$ and the cantilever stiffness $k$. In order to measure $F$ one has to compute the work $W_\tau$ performed by $F$ in the time $\tau$ after each rise. Since $F$ is constant, 
\begin{eqnarray}
W_\tau=F \int_0^\tau \dot x \ dt= F \ \Delta X _\tau
\label{eq:work}
\end{eqnarray}
where $\Delta X_\tau=x_f(\tau)-x_i(0)$ is the difference between the final value $x_f$ at time $\tau$ and the initial value $x_i$ just before the rise of $\Vs$. Since the protocol is equivalent when the applied force goes up or down, the analysis also uses both directions to accumulate more data. $F$ being constant, we can write that $\mathcal{P}(W_\tau)\propto \mathcal{P}(\Delta X_\tau)$. Thus eq.~\ref{eq:TFT} can be rewritten as:
\begin{equation}
\Phi (\Delta X_\tau) = \frac{F}{k_B T}\Delta X_\tau, \ \ \ \forall \tau
\label{eq:TFT_X}
\end{equation} 
where the symmetry function $\Phi$ is defined as: 
\begin{equation}
\Phi (\Delta X_\tau) = \ln \left( \frac{\mathcal{P}(\Delta X_\tau ) }{\mathcal{P}(- \Delta X_\tau )}\right)
\end{equation}
In eq.~\ref{eq:TFT_X} the only unknown is $F$ which can be determined by a linear fit of $\Phi (\Delta X_\tau)$ versus $\Delta X_\tau$. The symmetry function $\Phi$ can be easily determined by measuring for each rise of the square wave $\Delta X_\tau$ and by computing its pdf $\mathcal{P}(\Delta X{_\tau} )$. An example of the resulting distribution, taken over a full experiment is shown in fig.~\ref{fig:Gaussienne}.

\begin{figure}[tb]
	\centering
	\includegraphics[width =\linewidth]{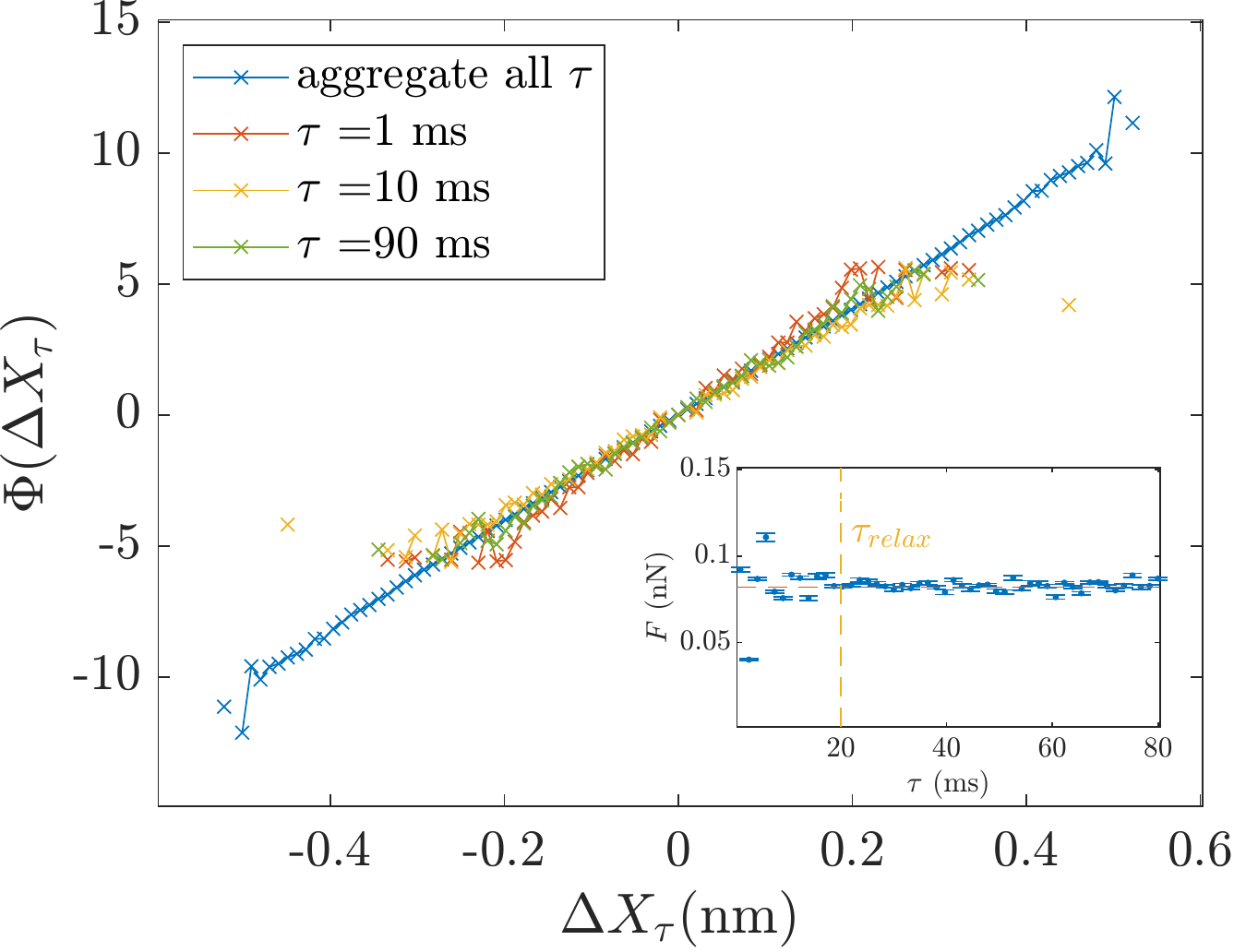}
	\caption{{The symmetry function $\Phi(\Delta X_\tau)$ at different times $\tau$: aggregate of all $\tau$ (blue),  \SI{1}{\milli\second} (red), \SI{10}{\milli\second} (yellow), \SI{100}{\milli\second} (green). As expected from the TFT (eq.~\ref{eq:TFT_X}), $\Phi$ is linear in $\Delta X_\tau$. Its slope is $F/k_BT$, and as shown in the inset the value of $F$ deduced is independent on $\tau$, even if the system has not relaxed to equilibrium.}}
	\label{fig:SymmetryFunction}
\end{figure}

The corresponding $\Phi(\Delta X_\tau)$ is plotted in fig.~\ref{fig:SymmetryFunction} as a function of $\Delta X_\tau$ and the slope 
of the linear fit is a measure of $F/k_B T$. We present three different values of $\tau$: $\SI{1}{\milli\second}${, right after the force jump}, $\SI{10}{\milli\second}$, during the relaxation of the cantilever and $\SI{100}{\milli\second}$, in the equilibrium plateau. As expected from the TFT, $\Phi(\Delta X_\tau)$ is independent of the value of $\tau$. The important point here is that, once the temperature $T$ is known, this force measurement based on TFT is independent of any calibration of the device (except $x$) and on the viscous dissipation. It also allows us to recover the value of the stiffness $k$, the distance $d$ and the contact potential $V_c$ as we show in the following.

It has to be pointed out that, as $\Pdx$ is Gaussian (see fig.~\ref{fig:Gaussienne}) the function $\Phi(\Delta X_\tau)$ in eq.~\ref{eq:TFT_X} takes a simple form: 
\begin{equation}
\Phi(\Delta X_\tau) = 2\frac{\langle \Delta X_\tau \rangle}{\stdX^2} \Delta X_\tau
\label{eq:Phigau}
\end{equation}
where $\langle \Delta X_\tau \rangle$ is the mean value of $\Delta X_\tau$ and $\stdX$ its standard deviation. From eq.~\ref{eq:TFT_X} and eq.~\ref{eq:Phigau} we get:
\begin{equation}
F = 2 k_B T \frac{\langle \Delta X_\tau \rangle}{\stdX^2}
\label{eq:Fgaus}
\end{equation}
For the linear fit of $\Phi$ and in eq \ref{eq:Phigau} and \ref{eq:Fgaus}, the value of $ \langle \Delta X_\tau \rangle$ and $\stdX$ are computed using the values for all $\tau$ and thus do not depend on $\tau$ anymore.

This allows for fast force measurements, because the estimation of $F$ using eq.~\ref{eq:Fgaus} is less affected than the linear fit (fig.~\ref{fig:SymmetryFunction}) by the low statistics on the values of $\Phi(\Delta X_\tau)$ far from the mean $\langle \Delta X_\tau \rangle$. 
The two methods (the linear fit and the Gaussian approximation) give the same results. 

\begin{figure}[tb]
	\centering
	\includegraphics[width= \linewidth]{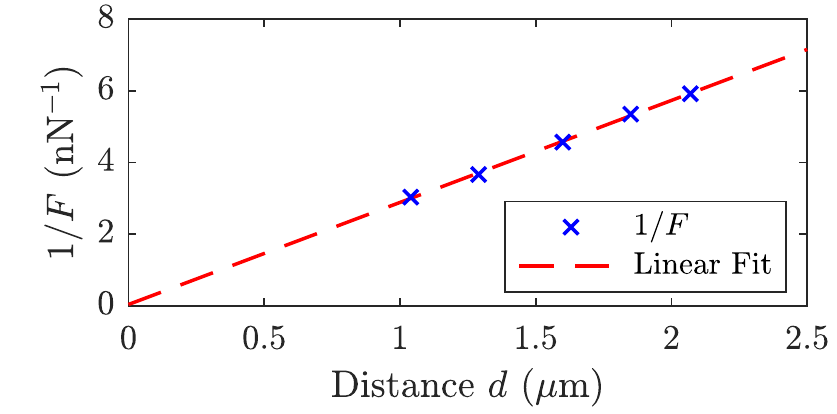}
	\caption{The inverse of the force $1/F$ measured with the TFT for an applied voltage $V_{sq} = \SI{250}{\milli\volt}$ at distances $d$ between $\SI{1}{\micro\meter}$ and $\SI{2}{\micro\meter}$, and linear fit. The linear trend expected from equation \ref{eq:elecforce} is recovered. The origin of the horizontal axis is chosen so that $1/F =0$ when $d=0$.}
	\label{fig:ForceVsDistance}
\end{figure}

To estimate the dependence on $d$ of the force $F$, measured using the TFT for an applied voltage $V_{sq} = \SI{250}{\milli\volt}$, we repeat the measure at different distances by displacing the plane with the piezo. {The results are shown in fig.~\ref{fig:ForceVsDistance} where we plot $1/F$ versus the distance $d$ controlled by piezo. We define the origin of distance such that $d=0$ when $1/F=0$. $d$ calibrated in this way now reflects the real sphere / plane distance, with an uncertainty of $\pm\SI{40}{nm}$ estimated from the linear fit at distances larger than 1 $\mu$m which are very safe to avoid surfaces damage. We recover the trend $ \Fs \propto 1/d$ as expected from equation \ref{eq:elecforce}.}

\begin{figure}[tb]
	\centering
	\includegraphics[width=\linewidth]{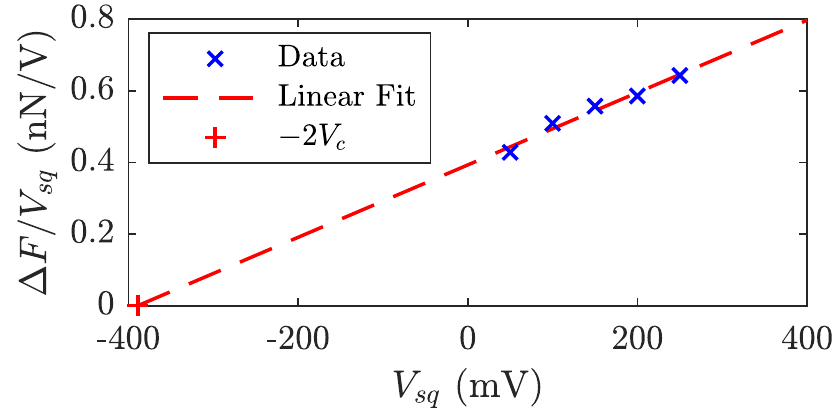}
	\caption{The ratio $\Delta F/\Vs$ as a function of $\Vs$ and the estimation of the contact potential $V_c$.}
	\label{fig:RegPotentielContact}
\end{figure}

Let us now estimate the contact potential by using the quadratic dependence of $F$ in $V_c$:
\begin{equation}
F = A [ (\Vs+V_c)^2 - V_c^2] = A (\Vs + 2 V_c) \Vs
\end{equation}
with $A=\pi \epsilon_0 R/d$ according to eq.~\ref{eq:elecforce}. Thus one can obtain $V_c$ and $A$ by doing a linear fit of the function $F/\Vs$ versus $\Vs$, whose values, measured at $d=\SI{1.95}{\micro\meter}$, are plotted in fig.~\ref{fig:RegPotentielContact}. The fit gives $V_c=\SI{194 \pm 15}{\milli\volt}$, and a slope $A=\SI{1.01 \pm 0.08e-9}{\newton\per\volt^2} $. The expected value for $A$ at $d=\SI{1.95}{\micro\meter}$ and $R=\SI{76}{\micro\meter}$ is $\SI{1.08e-9}{\newton\per\volt^2} $, in good agreement with the measured one.

{Furthermore, as $V_c$ is known, the measurements of $F$ as a function of $d$ plotted in fig.~\ref{fig:ForceVsDistance} can be used to measure the prefactor $B=\pi \epsilon_0 R (V^2-V_c^2)$ in eq.~\ref{eq:elecforce} where $V=\Vs+V_c$. The slope of $1/F$ versus $d$ in fig.~\ref{fig:ForceVsDistance} is $B{^{-1}}=\SI{2.85\pm 0.09e15}{\newton^{-1}\meter^{-1}}$ in good agreement again with the expected value of $\SI{2.97\pm 0.14e15}{\newton^{-1}\meter^{-1}}$ (where the uncertainty comes from that on $V_c$).}

It is important to stress again that all these results have been obtained without knowing the stiffness of the cantilever and without making any model of the cantilever dynamics and of the contribution of the high order modes to the measurements. However from the previous measurement the stiffness can be measured using TFT as 
\begin{equation}
k = F/ \langle \Delta X \rangle \label{eq:TFT_k}
\end{equation}
where $F$ is measured using the slope of $\Phi(\Delta X_\tau)$ and eq.~\ref{eq:TFT_X}. Another estimation of $k$ can be computed using the Gaussian approximation (eq.~\ref{eq:Fgaus}):
\begin{equation}
k = \frac{2 k_B T }{\stdX^2} \label{eq:kGaus}
\end{equation}
The values resulting from the measurements at different voltages and obtained by using the linear fit (eqs. \ref{eq:TFT_X} and \ref{eq:TFT_k}) and the Gaussian approximation (eqs. \ref{eq:Fgaus} and \ref{eq:kGaus}) are shown in table \ref{Tab:Valeursk}.

\begin{table}[tb]
	{\begin{tabular}{ |c|c|c|c|c|}
		\hline
		$\Vs$ (mV) & $F$ (pN) & $F_\mathrm{Gauss}$ (pN) & $k$ ($\si{\newton/\meter}$) & $k_\mathrm{Gauss}$ ($\si{\newton/\meter}$) \\
		 &from eq.~\ref{eq:TFT_X} & from eq.~\ref{eq:TFT_k} & from eq.~\ref{eq:Fgaus} & from eq.~\ref{eq:kGaus} \\
		\hline
		$50 $ & $21.4$ & $21.4$ & $0.398$ & $0.398$ \\
		$100$ & $51.0$ & $50.9$ & $0.400$ & $0.399$ \\
		$150$ & $84.3$ & $83.6$ & $0.400$ & $0.396$ \\
		$200$ & $122.8$ & $117.1$ & $0.413$ & $0.394$ \\
		$250$ & $154.9$ & $160.6$ & $0.388$ & $0.402$ \\
		\hline
	\end{tabular}}
	\caption{Values of the force and stiffness obtained using the TFT with and without the Gaussian approximation. This gives an average value across the experiments of $k = \SI{0.400\pm0.008}{\newton/\meter}$ and $k_\mathrm{Gauss} = \SI{0.398\pm0.003}{\newton/\meter}$.}
	\label{Tab:Valeursk}
\end{table}

 \begin{figure}[t]
 	\centering
 	\includegraphics[width=\linewidth]{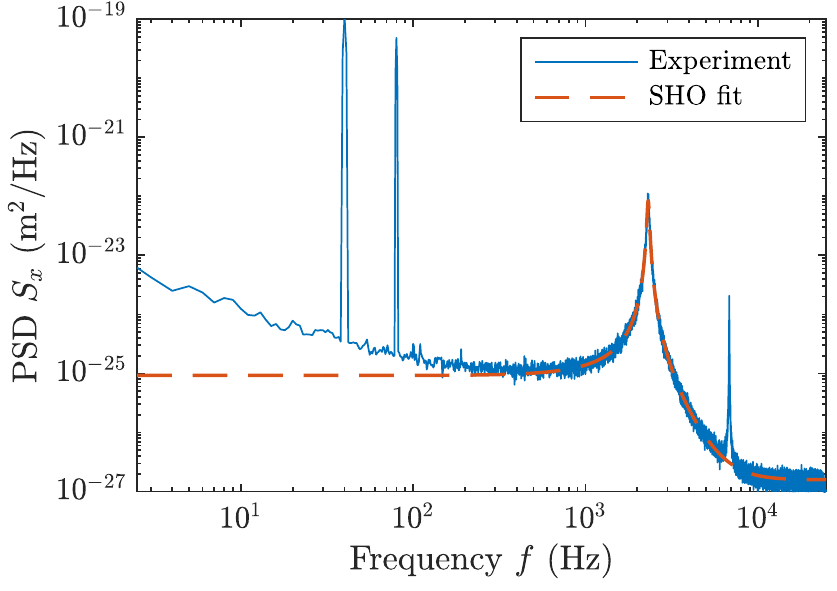}
 	\caption{The power spectrum density $S_x$ of the deflection when the cantilever is subjected to an electrostatic driving at $\omega_d = 2 \pi\ 40 \ \si{\radian /\second}$. The quadratic dependency in voltage is evidence by the narrow peaks at $\omega_d$ (term $V_0V_c$ in eq.~\ref{SpectreElec}) and $2\omega_d$ (term in $V_0^2$). The rest of the spectrum is only thermal noise driven. The SHO fit is performed on a $\SI{1}{k\hertz}$ window around the first resonance of the cantilever. The second resonance peak corresponds to the first torsion mode. The $1/f$ noise at low frequency is due to the viscoelasticity of the gold coating and is not taken into account in the SHO model \cite{Paolino-2009-Nanotech}.} 
 	\label{fig:SpectreSHO}
 \end{figure}
 
In spite of the fact that the TFT gives us an accurate and precise description of the interaction we check the results using standard calibration techniques. The AFM cantilever is described by a simple harmonic oscillator (SHO) of stiffness $k$, mass $m$, and viscous damping coefficient $\gamma$. Its transfer function is: 
\begin{equation} \label{eq:G}
G(\omega) =\frac{\tilde x(\omega)}{\tilde F(\omega)}=\frac{1}{k-m \omega^2 + i \gamma \omega}
\end{equation}
where the tilde designs Fourier transform and $\omega=2 \pi f$ the angular frequency. The thermal noise power spectrum density in deflection $S_x$ of such an SHO is 
\begin{equation}
S_x(\omega) = \frac{2 k_B T \gamma}{\pi} \frac{1}{(k-m \omega^2)^2 + \gamma^2 \omega^2}
\end{equation}
The value of the stiffness $k = \SI{.40\pm0.02}{\newton/\meter}$ extracted from the SHO fit of the experimental data of fig. \ref{fig:SpectreSHO} can be cross-checked by the direct measure of the variance $x(t)$ which is related to $k$ by the energy equipartition: $k = k_B T/ \langle x^2 \rangle$. Both methods give a stiffness equal within error bars to the one estimated using TFT.

To measure the distance and contact potential, we can use a technique derived for Kelvin Probe Force Microscopy, applying a voltage $V_d= V_0 \cos(\omega_d t)$ to the cantilever. The force in the Fourier Space is:
\begin{multline}
 \tilde F(\omega) = \frac{\pi \epsilon_0 R}{d} \left[ (V_c^2 + \frac{V_0^2}{2}) \delta(\omega) + 2 V_0 V_c \delta(\omega - \omega_d) \right. \\ \left. + \frac{V_0^2}{2} \delta( \omega -2\omega_d) \right] \label{SpectreElec}
\end{multline}
The psd of $x$ in presence of the electric forcing is plotted infig.  \ref{fig:SpectreSHO}, showing two peaks at $\omega_d$ and $2\omega_d$. The response at $2\omega_d$ is only caused by the applied voltage and allows for a measurement of the distance, whereas the term at $\omega_d$ couples the applied voltage to the contact potential and allows us to measure $V_c$ \cite{lecunuder2018,de2009no}.

Using eqs.~\ref{eq:G} and \ref{SpectreElec} at $\omega = 2\omega_d$, we have:
\begin{align}
 \tilde{x}(2\omega_d) & = G(2\omega_d) \frac{\pi \epsilon_0 R}{d} \frac{V_0^2}{2} \\
 d &= G(2\omega_d) \pi \epsilon_0 R \frac{\widetilde{V^2}(2\omega_d)}{\tilde{x}(2\omega_d)} 
\end{align}
The term $\widetilde{V^2}(2\omega_d)/\tilde{x}(2\omega_d)$ is numerically computed as the transfer function between $V^2(t)$ and $x(t)$.
This method is used to measure the distances during the experiments in good agreement with the values measured with TFT.

Using eqs.~\ref{eq:G} and \ref{SpectreElec} at $\omega = \omega_d$, we have:
\begin{align}
\tilde{x} (\omega_d) & = G(\omega_d) \frac{\pi \epsilon_0 R}{d} 2 V_0 V_c \\
V_c & = \frac{1}{G(\omega_d)}\frac{d}{2\pi \epsilon_0 R} \frac{\tilde{x}(\omega_d)}{\tilde{V} (\omega_d)} \\
V_c &= \frac{1}{2} \frac{G(2 \omega_d)}{G(\omega_d)} \frac{\widetilde{V^2}(2\omega_d)}{\tilde{x}(2\omega_d)} \frac{\tilde{x}(\omega_d)}{\tilde{V}(\omega_d)}
\end{align}
The term $\tilde{x}(\omega_d)/\tilde{V} (\omega_d)$ is obtained as the transfer function between the measured deflection $x(t)$ and the applied voltage $V(t)$. This gives a reference value for the contact potential of $V_c = \SI{207\pm5}{\milli\volt}$, again in very good agreement with the value measured by TFT.

As a conclusion, we have shown that the TFT is a useful tool to check the accuracy of a force measurement by an AFM cantilever. It is independent on the viscosity and on the stiffness calibration of the system. The results are in perfect agreement with those of other methods with a comparable accuracy. Other techniques inspired by stochastic thermodynamics could be used but they are slightly more complex and we presented the TFT as a proof of principle experiment. 

The proposed method can be applied as soon as the measurement of the deflection $x$ is calibrated. This is straightforward for an interferometric AFM as we have used, as long as the measurement laser is carefully tuned at the tip position. For more common AFM setups and unknown tip-sample interaction, a calibration of the sensitivity of the detector (usually a 4 quadrant photodiode) should be performed separately, for example using a force curve on a hard surface.

Colloidal probes, as the one used in this letter, are commonly used in many applications, such as chemical sensing and detection, intermolecular or adhesion forces measurement\cite{Novascan}, elasticity characterisation in biology or soft matter, magnetic detection, etc. In this very well defined sphere plane geometry, for a known interaction at $d\ll R$ it is possible to extract also the sensitivity if $R$ is known. One has to use the force versus distance and force versus potential measurements (see figs. \ref{fig:ForceVsDistance}, \ref{fig:RegPotentielContact}) and eq.\ref{eq:elecforce} where the only unknown is the $x$ calibration. Another calibration possibility for colloidal probe is to use the viscous drag in a fluid of viscosity $\eta$, with $F_H = 6 \pi \eta R^2 \dot d /d$ for $d\ll R$, where in this case $d$ is the modulated control parameter instead of $V$. The analysis can be performed as presented here by computing the work of $F_H$~\cite{Brenner-1961}. Thus for such AFM probes using TFT we can estimate in the same measurement the calibration factor, the value of the force, the stiffness and the sphere/plane distance.

As a final remark let us emphasis that the technique of modulating $d$ can be also useful for estimating any kind of tip-sample interaction force in practical cases. Indeed, as long as the force can be considered constant on the explored deflection range, applying steps in $d$ faster that the response of the cantilever will result in steps in $F$. The analysis can then be performed with the TFT as presented here, leading to the force-distance determination.

{\bf Data Availability} Data available on request from the authors.

{\bf Aknowledments} The interaction with Baptiste Ferrero and Vincent Dolique is gratefully acknowledged.

\bigskip

\bibliography{ArticleTF_references}

\begin{thebibliography}{16}%
\makeatletter
\providecommand \@ifxundefined [1]{%
 \@ifx{#1\undefined}
}%
\providecommand \@ifnum [1]{%
 \ifnum #1\expandafter \@firstoftwo
 \else \expandafter \@secondoftwo
 \fi
}%
\providecommand \@ifx [1]{%
 \ifx #1\expandafter \@firstoftwo
 \else \expandafter \@secondoftwo
 \fi
}%
\providecommand \natexlab [1]{#1}%
\providecommand \enquote  [1]{``#1''}%
\providecommand \bibnamefont  [1]{#1}%
\providecommand \bibfnamefont [1]{#1}%
\providecommand \citenamefont [1]{#1}%
\providecommand \href@noop [0]{\@secondoftwo}%
\providecommand \href [0]{\begingroup \@sanitize@url \@href}%
\providecommand \@href[1]{\@@startlink{#1}\@@href}%
\providecommand \@@href[1]{\endgroup#1\@@endlink}%
\providecommand \@sanitize@url [0]{\catcode `\\12\catcode `\$12\catcode
  `\&12\catcode `\#12\catcode `\^12\catcode `\_12\catcode `\%12\relax}%
\providecommand \@@startlink[1]{}%
\providecommand \@@endlink[0]{}%
\providecommand \url  [0]{\begingroup\@sanitize@url \@url }%
\providecommand \@url [1]{\endgroup\@href {#1}{\urlprefix }}%
\providecommand \urlprefix  [0]{URL }%
\providecommand \Eprint [0]{\href }%
\providecommand \doibase [0]{http://dx.doi.org/}%
\providecommand \selectlanguage [0]{\@gobble}%
\providecommand \bibinfo  [0]{\@secondoftwo}%
\providecommand \bibfield  [0]{\@secondoftwo}%
\providecommand \translation [1]{[#1]}%
\providecommand \BibitemOpen [0]{}%
\providecommand \bibitemStop [0]{}%
\providecommand \bibitemNoStop [0]{.\EOS\space}%
\providecommand \EOS [0]{\spacefactor3000\relax}%
\providecommand \BibitemShut  [1]{\csname bibitem#1\endcsname}%
\let\auto@bib@innerbib\@empty
\bibitem [{\citenamefont {Fronczak}\ \emph {et~al.}(2018)\citenamefont
  {Fronczak}, \citenamefont {Browne}, \citenamefont {Krenek}, \citenamefont
  {Beaudoin},\ and\ \citenamefont {Corti}}]{Calib_1}%
  \BibitemOpen
  \bibfield  {author} {\bibinfo {author} {\bibfnamefont {S.~G.}\ \bibnamefont
  {Fronczak}}, \bibinfo {author} {\bibfnamefont {C.~A.}\ \bibnamefont
  {Browne}}, \bibinfo {author} {\bibfnamefont {E.~C.}\ \bibnamefont {Krenek}},
  \bibinfo {author} {\bibfnamefont {S.~P.}\ \bibnamefont {Beaudoin}}, \ and\
  \bibinfo {author} {\bibfnamefont {D.~S.}\ \bibnamefont {Corti}},\ }\href
  {\doibase 10.1016/j.jcis.2018.01.108} {\bibfield  {journal} {\bibinfo
  {journal} {J. Colloid Interface Sci.}\ }\textbf {\bibinfo {volume} {517}},\
  \bibinfo {pages} {213} (\bibinfo {year} {2018})}\BibitemShut {NoStop}%
\bibitem [{\citenamefont {Payam}\ \emph {et~al.}(2018)\citenamefont {Payam},
  \citenamefont {Trewby},\ and\ \citenamefont {Voitchovsky}}]{Calib_2}%
  \BibitemOpen
  \bibfield  {author} {\bibinfo {author} {\bibfnamefont {A.~F.}\ \bibnamefont
  {Payam}}, \bibinfo {author} {\bibfnamefont {W.}~\bibnamefont {Trewby}}, \
  and\ \bibinfo {author} {\bibfnamefont {K.}~\bibnamefont {Voitchovsky}},\
  }\href {\doibase 10.1063/1.5009071} {\bibfield  {journal} {\bibinfo
  {journal} {Appl. Phys. Lett.}\ }\textbf {\bibinfo {volume} {112}} (\bibinfo
  {year} {2018}),\ 10.1063/1.5009071}\BibitemShut {NoStop}%
\bibitem [{\citenamefont {Dagdeviren}\ \emph {et~al.}(2019)\citenamefont
  {Dagdeviren}, \citenamefont {Miyahara}, \citenamefont {Mascaro},\ and\
  \citenamefont {Grutter}}]{Calib_3}%
  \BibitemOpen
  \bibfield  {author} {\bibinfo {author} {\bibfnamefont {O.~E.}\ \bibnamefont
  {Dagdeviren}}, \bibinfo {author} {\bibfnamefont {Y.}~\bibnamefont
  {Miyahara}}, \bibinfo {author} {\bibfnamefont {A.}~\bibnamefont {Mascaro}}, \
  and\ \bibinfo {author} {\bibfnamefont {P.}~\bibnamefont {Grutter}},\
  }\href@noop {} {\bibfield  {journal} {\bibinfo  {journal} {Rev. Sci. Instr.}\
  }\textbf {\bibinfo {volume} {90}} (\bibinfo {year} {2019})}\BibitemShut
  {NoStop}%
\bibitem [{\citenamefont {Laurent}\ \emph {et~al.}(2012)\citenamefont
  {Laurent}, \citenamefont {Sellier}, \citenamefont {Mosset}, \citenamefont
  {Huant},\ and\ \citenamefont {Chevrier}}]{Laurent_2012}%
  \BibitemOpen
  \bibfield  {author} {\bibinfo {author} {\bibfnamefont {J.}~\bibnamefont
  {Laurent}}, \bibinfo {author} {\bibfnamefont {H.}~\bibnamefont {Sellier}},
  \bibinfo {author} {\bibfnamefont {A.}~\bibnamefont {Mosset}}, \bibinfo
  {author} {\bibfnamefont {S.}~\bibnamefont {Huant}}, \ and\ \bibinfo {author}
  {\bibfnamefont {J.}~\bibnamefont {Chevrier}},\ }\href@noop {} {\bibfield
  {journal} {\bibinfo  {journal} {Phys. Rev. B}\ }\textbf {\bibinfo {volume}
  {85}},\ \bibinfo {pages} {035426} (\bibinfo {year} {2012})}\BibitemShut
  {NoStop}%
\bibitem [{\citenamefont {Paolino}\ \emph {et~al.}(2013)\citenamefont
  {Paolino}, \citenamefont {Aguilar~Sandoval},\ and\ \citenamefont
  {Bellon}}]{paolino2013RSI}%
  \BibitemOpen
  \bibfield  {author} {\bibinfo {author} {\bibfnamefont {P.}~\bibnamefont
  {Paolino}}, \bibinfo {author} {\bibfnamefont {F.}~\bibnamefont
  {Aguilar~Sandoval}}, \ and\ \bibinfo {author} {\bibfnamefont
  {L.}~\bibnamefont {Bellon}},\ }\href@noop {} {\bibfield  {journal} {\bibinfo
  {journal} {Rev. Sci. Instrum.}\ }\textbf {\bibinfo {volume} {84}},\ \bibinfo
  {pages} {095001} (\bibinfo {year} {2013})}\BibitemShut {NoStop}%
\bibitem [{\citenamefont {De~Man}\ \emph {et~al.}(2009)\citenamefont {De~Man},
  \citenamefont {Heeck},\ and\ \citenamefont {Iannuzzi}}]{de2009no}%
  \BibitemOpen
  \bibfield  {author} {\bibinfo {author} {\bibfnamefont {S.}~\bibnamefont
  {De~Man}}, \bibinfo {author} {\bibfnamefont {K.}~\bibnamefont {Heeck}}, \
  and\ \bibinfo {author} {\bibfnamefont {D.}~\bibnamefont {Iannuzzi}},\
  }\href@noop {} {\bibfield  {journal} {\bibinfo  {journal} {Phys. Rev. A}\
  }\textbf {\bibinfo {volume} {79}},\ \bibinfo {pages} {024102} (\bibinfo
  {year} {2009})}\BibitemShut {NoStop}%
\bibitem [{\citenamefont {Bellon}(2008)}]{Bellon_2008}%
  \BibitemOpen
  \bibfield  {author} {\bibinfo {author} {\bibfnamefont {L.}~\bibnamefont
  {Bellon}},\ }\href@noop {} {\bibfield  {journal} {\bibinfo  {journal} {J.
  Appl. Phys.}\ }\textbf {\bibinfo {volume} {104}},\ \bibinfo {pages} {104906}
  (\bibinfo {year} {2008})}\BibitemShut {NoStop}%
\bibitem [{\citenamefont {Ciliberto}(2017)}]{Ciliberto_PRX}%
  \BibitemOpen
  \bibfield  {author} {\bibinfo {author} {\bibfnamefont {S.}~\bibnamefont
  {Ciliberto}},\ }\href@noop {} {\bibfield  {journal} {\bibinfo  {journal}
  {Phys. Rev. X}\ }\textbf {\bibinfo {volume} {7}},\ \bibinfo {pages} {021051}
  (\bibinfo {year} {2017})}\BibitemShut {NoStop}%
\bibitem [{\citenamefont {Seifert}(2012)}]{Seifert_2012}%
  \BibitemOpen
  \bibfield  {author} {\bibinfo {author} {\bibfnamefont {U.}~\bibnamefont
  {Seifert}},\ }\href {\doibase 10.1088/0034-4885/75/12/126001} {\bibfield
  {journal} {\bibinfo  {journal} {Rep. Prog. Phys.}\ }\textbf {\bibinfo
  {volume} {75}},\ \bibinfo {pages} {126001} (\bibinfo {year}
  {2012})}\BibitemShut {NoStop}%
\bibitem [{\citenamefont {Evans}\ \emph {et~al.}(1993)\citenamefont {Evans},
  \citenamefont {Cohen},\ and\ \citenamefont {Morriss}}]{EvansFT}%
  \BibitemOpen
  \bibfield  {author} {\bibinfo {author} {\bibfnamefont {D.~J.}\ \bibnamefont
  {Evans}}, \bibinfo {author} {\bibfnamefont {E.~G.~D.}\ \bibnamefont {Cohen}},
  \ and\ \bibinfo {author} {\bibfnamefont {G.~P.}\ \bibnamefont {Morriss}},\
  }\href {\doibase 10.1103/PhysRevLett.71.2401} {\bibfield  {journal} {\bibinfo
   {journal} {Phys. Rev. Lett.}\ }\textbf {\bibinfo {volume} {71}},\ \bibinfo
  {pages} {2401} (\bibinfo {year} {1993})}\BibitemShut {NoStop}%
\bibitem [{\citenamefont {Le~Cunuder}\ \emph {et~al.}(2018)\citenamefont
  {Le~Cunuder}, \citenamefont {Petrosyan}, \citenamefont {Palasantzas},
  \citenamefont {Svetovoy},\ and\ \citenamefont {Ciliberto}}]{lecunuder2018}%
  \BibitemOpen
  \bibfield  {author} {\bibinfo {author} {\bibfnamefont {A.}~\bibnamefont
  {Le~Cunuder}}, \bibinfo {author} {\bibfnamefont {A.}~\bibnamefont
  {Petrosyan}}, \bibinfo {author} {\bibfnamefont {G.}~\bibnamefont
  {Palasantzas}}, \bibinfo {author} {\bibfnamefont {V.}~\bibnamefont
  {Svetovoy}}, \ and\ \bibinfo {author} {\bibfnamefont {S.}~\bibnamefont
  {Ciliberto}},\ }\href@noop {} {\bibfield  {journal} {\bibinfo  {journal}
  {Phys. Rev. B}\ }\textbf {\bibinfo {volume} {98}},\ \bibinfo {pages}
  {201408(R)} (\bibinfo {year} {2018})}\BibitemShut {NoStop}%
\bibitem [{\citenamefont {Durand}(1964)}]{Durand_elect}%
  \BibitemOpen
  \bibfield  {author} {\bibinfo {author} {\bibfnamefont {E.}~\bibnamefont
  {Durand}},\ }\href@noop {} {\emph {\bibinfo {title} {Electrostatique, volume
  1}}}\ (\bibinfo  {publisher} {Masson},\ \bibinfo {year} {1964})\BibitemShut
  {NoStop}%
\bibitem [{\citenamefont {Crowley}(2008)}]{elec_sph_pla}%
  \BibitemOpen
  \bibfield  {author} {\bibinfo {author} {\bibfnamefont {J.~M.}\ \bibnamefont
  {Crowley}},\ }\href@noop {} {\bibfield  {journal} {\bibinfo  {journal} {Proc.
  ESA Annual Meeting on Electrostatics}\ ,\ \bibinfo {pages} {Paper D1}}
  (\bibinfo {year} {2008})}\BibitemShut {NoStop}%
\bibitem [{\citenamefont {Paolino}\ and\ \citenamefont
  {Bellon}(2009)}]{Paolino-2009-Nanotech}%
  \BibitemOpen
  \bibfield  {author} {\bibinfo {author} {\bibfnamefont {P.}~\bibnamefont
  {Paolino}}\ and\ \bibinfo {author} {\bibfnamefont {L.}~\bibnamefont
  {Bellon}},\ }\href {\doibase 10.1088/0957-4484/20/40/405705} {\bibfield
  {journal} {\bibinfo  {journal} {Nanotechnology}\ }\textbf {\bibinfo {volume}
  {20}},\ \bibinfo {pages} {405705} (\bibinfo {year} {2009})}\BibitemShut
  {NoStop}%
\bibitem [{\citenamefont {Novascan}(2020)}]{Novascan}%
  \BibitemOpen
  \bibfield  {author} {\bibinfo {author} {\bibnamefont {Novascan}},\
  }\href@noop {} {\bibfield  {journal} {\bibinfo  {journal} {Novascan AFM
  Probes Brochure}\ } (\bibinfo {year} {2020})}\BibitemShut {NoStop}%
\bibitem [{\citenamefont {Brenner}(1961)}]{Brenner-1961}%
  \BibitemOpen
  \bibfield  {author} {\bibinfo {author} {\bibfnamefont {H.}~\bibnamefont
  {Brenner}},\ }\href {\doibase 10.1016/0009-2509(61)80035-3} {\bibfield
  {journal} {\bibinfo  {journal} {Chem. Eng. Sci.}\ }\textbf {\bibinfo {volume}
  {16}},\ \bibinfo {pages} {242 } (\bibinfo {year} {1961})}\BibitemShut
  {NoStop}%
\end{thebibliography}%

\end{document}